\documentclass[fleqn,usenatbib]{mnras}

\usepackage{newtxtext,newtxmath}

\usepackage[T1]{fontenc}

\DeclareRobustCommand{\VAN}[3]{#2}
\let\VANthebibliography\thebibliography
\def\thebibliography{\DeclareRobustCommand{\VAN}[3]{##3}\VANthebibliography}

\usepackage{graphicx}	
\usepackage{amsmath}	
\usepackage{bm}
\usepackage[normalem]{ulem}

\newcommand\oiii    	{$\mathrm{\left[ O \textsc{iii}\right] }$}
\newcommand\oi	    	{$\mathrm{\left[ O \textsc{i}\right] }$}
\newcommand\moi	    	{\mathrm{\left[ O \textsc{i}\right] }}
\newcommand\nii	    	{$\mathrm{\left[ N \textsc{ii}\right] }$}
\newcommand\mnii    	{\mathrm{\left[ N \textsc{ii}\right] }}
\newcommand\sii	    	{$\mathrm{\left[ S \textsc{ii}\right] }$}
\newcommand\msii    	{\mathrm{\left[ S \textsc{ii}\right] }}

\newcommand\sbar        {\textsc{Sbar}}
\newcommand\wbar        {\textsc{Wbar}}
\newcommand\ubar        {\textsc{Ubar}}

\newcommand\mfagnsbar    {f_{\mathrm{AGN, \sbar}}}
\newcommand\mfagnwbar    {f_{\mathrm{AGN, \wbar}}}
\newcommand\mfagnubar    {f_{\mathrm{AGN, \ubar}}}

\newcommand\mmsun	      {\rm{M}_{\odot}}
\newcommand\ha          {H$\upalpha$}

\newcommand\hb          {H$\upbeta$}
\newcommand\mstar       {$M_{\ast}$}
\newcommand\mmstar       {M_{\ast}}

\title[GZ DESI: Bars and AGN]{Galaxy Zoo DESI: large-scale bars as a secular mechanism for triggering AGN}

\author[I. L. Garland et al.]{
Izzy L. Garland,$^{1}$\thanks{E-mail: i.garland@lancaster.ac.uk}
Mike Walmsley,$^{2,3}$
Maddie S. Silcock,$^{1,4}$
Leah M. Potts,$^{1}$
Josh Smith,$^{1}$ \newauthor
Brooke D. Simmons,$^{1}$
Chris J. Lintott,$^{5}$
Rebecca J. Smethurst,$^{5}$
James M. Dawson,$^{6,7}$
William C. Keel,$^{8}$\newauthor
Sandor Kruk,$^{9}$
Kameswara Bharadwaj Mantha,$^{10,11}$
Karen L. Masters,$^{12}$
David O'Ryan,$^{1}$
J\"{u}rgen J. Popp,$^{13}$\newauthor
Matthew R. Thorne$^{1}$
\\
$^{1}$Physics Department, Lancaster University, Lancaster, LA1 4YB, UK \\
$^{2}$Dunlap Institute for Astronomy and Astrophysics, University of Toronto, 50 St. George Street, Toronto, ON M5S 3H4, Canada \\
$^{3}$Jodrell Bank Centre for Astrophysics, Department of Physics \& Astronomy, University of Manchester, Oxford Road, Manchester, M13 9PL, UK \\
$^{4}$Centre for Astrophysics Research, University of Hertfordshire, College Lane, Hatfield, AL10 9AB, UK \\
$^{5}$Oxford Astrophysics, Department of Physics, University of Oxford, Denys Wilkinson Building, Keble Road, Oxford, OX1 3RH, UK \\
$^{6}$South African Radio Astronomy Observatory (SARAO), Black River Park North, 2 Fir St, Cape Town, South Africa, 7925\\
$^{7}$Department of Physics and Electronics, Rhodes University, PO Box 94, Makhanda 6140, South Africa\\
$^{8}$Department of Physics and Astronomy, University of Alabama, 206 Gallalee Hall, 514 University Blvd. Tuscaloosa, AL 35487-0324, USA\\
$^{9}$European Space Agency (ESA), European Space Astronomy Centre (ESAC), Camino Bajo del Castillo s/n, 28692, Villaneuva de la Ca\~nada, Madrid, Spain\\
$^{10}$School of Physics and Astronomy, University of Minnesota, Minneapolis, Minnesota, 55455, USA\\
$^{11}$Minnesota Institute for Astrophysics, University of Minnesota, Minneapolis, Minnesota, 55455, USA\\
$^{12}$Departments of Physics and Astronomy, Haverford College, Lancaster Avenue, Ardmore, PA, 19041 USA \\
$^{13}$School of Physical Sciences, The Open University, Milton Keynes, MK7 6AA, UK \\
}

\date{Accepted XXX. Received YYY; in original form ZZZ}

\pubyear{2023}

\begin{document}
\label{firstpage}
\pagerange{\pageref{firstpage}--\pageref{lastpage}}
\maketitle

\begin{abstract}
Despite the evidence that supermassive black holes (SMBHs) co-evolve with their host galaxy, and that most of the growth of these SMBHs occurs via merger-free processes, the underlying mechanisms which drive this secular co-evolution are poorly understood.
We investigate the role that both strong and weak large-scale galactic bars play in mediating this relationship.
Using 72,940 disc galaxies in a volume-limited sample from Galaxy Zoo DESI, we analyse the active galactic nucleus (AGN) fraction in strongly barred, weakly barred, and unbarred galaxies up to $z = 0.1$ over a range of stellar masses and colours.
After controlling for stellar mass and colour, we find that the optically selected AGN fraction is $31.6 \pm 0.9$ per cent in strongly barred galaxies, $23.3 \pm 0.8$ per cent in weakly barred galaxies, and $14.2 \pm 0.6$ per cent in unbarred disc galaxies.
These are highly statistically robust results, strengthening the tantalising results in earlier works.
Strongly barred galaxies have a higher fraction of AGNs than weakly barred galaxies, which in turn have a higher fraction than unbarred galaxies.
Thus, while bars are not required in order to grow a SMBH in a disc galaxy, large-scale galactic bars appear to facilitate AGN fuelling, and the presence of a strong bar makes a disc galaxy more than twice as likely to host an AGN than an unbarred galaxy at all galaxy stellar masses and colours.
\end{abstract}

\begin{keywords}
galaxies: active -- galaxies: bar -- galaxies: disc -- galaxies: Seyfert
\end{keywords}



\section{Introduction}
Supermassive black holes (SMBHs) reside in the centre of the majority of galaxies, gaining most of their mass during active phases, where the accretion systems are known as active galactic nuclei (AGNs).
Yet what triggers the ``switch on'' of an AGN is equivocal.
This question is critical to understanding the interplay between AGNs and their host galaxies, including the effectiveness of AGN feedback and SMBH-galaxy co-evolution \citep[see e.g.,][for a review]{kormendy2013, heckman2014}.

Recent simulation studies have shown that the majority of SMBH growth occurs via secular (merger-free) mechanisms \citep{martin2018, mcalpine2020, smethurst2023}, meaning that mergers are not the primary drivers of the relationships known to exist between SMBHs and their host galaxies.
Disc-dominated, bulgeless galaxies have had a history free from major mergers (1:10 mass ratio) since at least $z\sim2$ \citep{martig2012}, and so by exclusively looking at a population of disc-dominated, bulgeless galaxies and the kiloparsec scale structures within them (such as large-scale galactic bars), we can gain a better understanding of AGN triggering in the absence of major mergers.
The bulge present in some disc-dominated galaxies could be merger-formed, but it could also be formed through a number of other mechanisms, including minor mergers, and potentially bars. By looking at a population of disc-dominated galaxies as a whole, we can investigate structures such as bars across the entire disc-dominated galaxy population.

Large-scale strong bars are observed at optical wavelengths in the Sloan Digital Sky Survey \citep[SDSS;][]{york2000} in $29.4\pm0.5$ per cent of disc galaxies at redshift $0.01<z<0.06$ \citep{masters2011}, and when using either a deeper optical survey or one with better seeing, such as DECaLS, this increases to around 45 per cent when combining galaxies with either weak or strong bars \citep{geron2021}.
This distinction between strong and weak bars is important, despite their being on a continuum, since work has shown that they may have different formation mechanisms \citep[e.g.,][]{geron2023}, although separating out strong and weak bars consistently poses a challenge.
In general, a bar is classified as strong if it dominates the galaxy flux, and weak as containing a smaller fraction of the total flux \citep{nair2010}.
These bars can cause transfers of a disc's angular momentum, leading to gas being transported down to the central kiloparsec region \citep{friedli1993, athanassoula2003}, where it could be accreted onto a black hole.
Thus, by tracing these kiloparsec-scale structures, we can gain insight into the dynamics within a galaxy that facilitate the transfer of angular momentum, and hence the fuelling which gives rise to the AGN characteristics that we observe.

Simulations have shown that it is physically possible for bars to provide the necessary inflow of gas to match the accretion rates we see in AGNs \citep{sakamoto1996, maciejewski2002, regan2004, lin2013}, and this is mirrored in observational work by \citet{smethurst2021}.
Several other studies have pointed to either an increase in the bar fraction of AGN hosts compared to inactive galaxies, or an increase in AGN fraction in barred galaxies compared to unbarred \citep{knapen2000, laine2002, coelho2011, oh2012, galloway2015, alonso2018, silva-lima2022, garland2023}.
However, many of these previous studies have suffered from low statistical significance or sensitivity to methodology and selection effects.

There are also a number of studies finding no correlation \citep[e.g.,][]{cheung2013, goulding2017}.
Thus, in this work, we aim to revisit this correlation between large-scale bars and AGNs, using Galaxy Zoo DESI \citep{walmsley2023b} to obtain robust morphologies from deeper imaging, and observed emission lines from SDSS MPA-JHU DR7\footnote{Available at \url{https://www.mpa-garching.mpg.de/SDSS/DR7/}} to determine the activity category of the systems in our sample.

Section \ref{sec:data_collection} discusses our sample selection and classification.
We present our results in Section \ref{sec:results}, followed by our discussions and conclusions in Sections \ref{sec:discussion} and \ref{sec:conclusions}.
Throughout this work, we consider AGNs and LINERs (low-ionisation nuclear emission line regions) to be two distinct categories, rather than LINERs being a subset of AGN.
We use WMAP9 cosmology \citep{hinshaw2013}{, incorporated via \textsc{Astropy}, where we assume a flat universe, $H_{0} = 69.3\,\mathrm{km}\,\mathrm{s}^{-1}\,\mathrm{Mpc}^{-1}$ and $\Omega_{m} = 0.287$.

\section{Data Collation}\label{sec:data_collection}
In the subsections below, we describe the use of multiple surveys to obtain the data required for this study.
We collate a sample of disc-dominated galaxies (divided into strongly barred, weakly barred and unbarred) which are either AGN hosts, star-forming, or undetermined.

\subsection{Sample Selection}
Galaxy Zoo DESI \citep[GZD;][]{walmsley2023b} uses machine learning to identify the morphology of 8.7M galaxies in the DESI Legacy Imaging Surveys: DECaLS, MzLS and BASS, plus DES.
Given the improved seeing on DESI compared to SDSS, we can push reliable morphology classifications to higher redshifts.
Full details of the methodology can be found in the release paper, and we summarise briefly here.

Given the size of the DESI Legacy Imaging Surveys, it was not feasible to collect morphological classifications from volunteers alone (such as in Galaxy Zoo 2), as this would take around 200 years at current classification rates.
Thus more efficient techniques are required.
\citet{walmsley2023b} trained deep learning models \citep{walmsley2023a} on 10M Galaxy Zoo volunteer votes over 401k galaxies from the DESI Legacy Surveys to classify galaxy morphology based on this training data.
Their models can typically predict what fraction of volunteers would give a particular answer to each question to within a mean vote fraction error of 10 per cent.

We match GZD within a 3'' radius to the MPA-JHU SDSS DR7 catalogue \citep[to obtain stellar masses, \mstar, colours and emission line fluxes;][]{kauffmann2003, salim2007} and the NYU-VAGC catalogue \citep[to obtain $k$-corrections;][]{blanton2005}, resulting in 793,824 galaxies.
Fig. \ref{fig:mag_z_vollim} shows absolute $r$-band magnitude versus redshift for the entire sample, as well as the volume-limited disc galaxy sample (described below).

\begin{figure}
    \centering
	\includegraphics[width=\columnwidth]{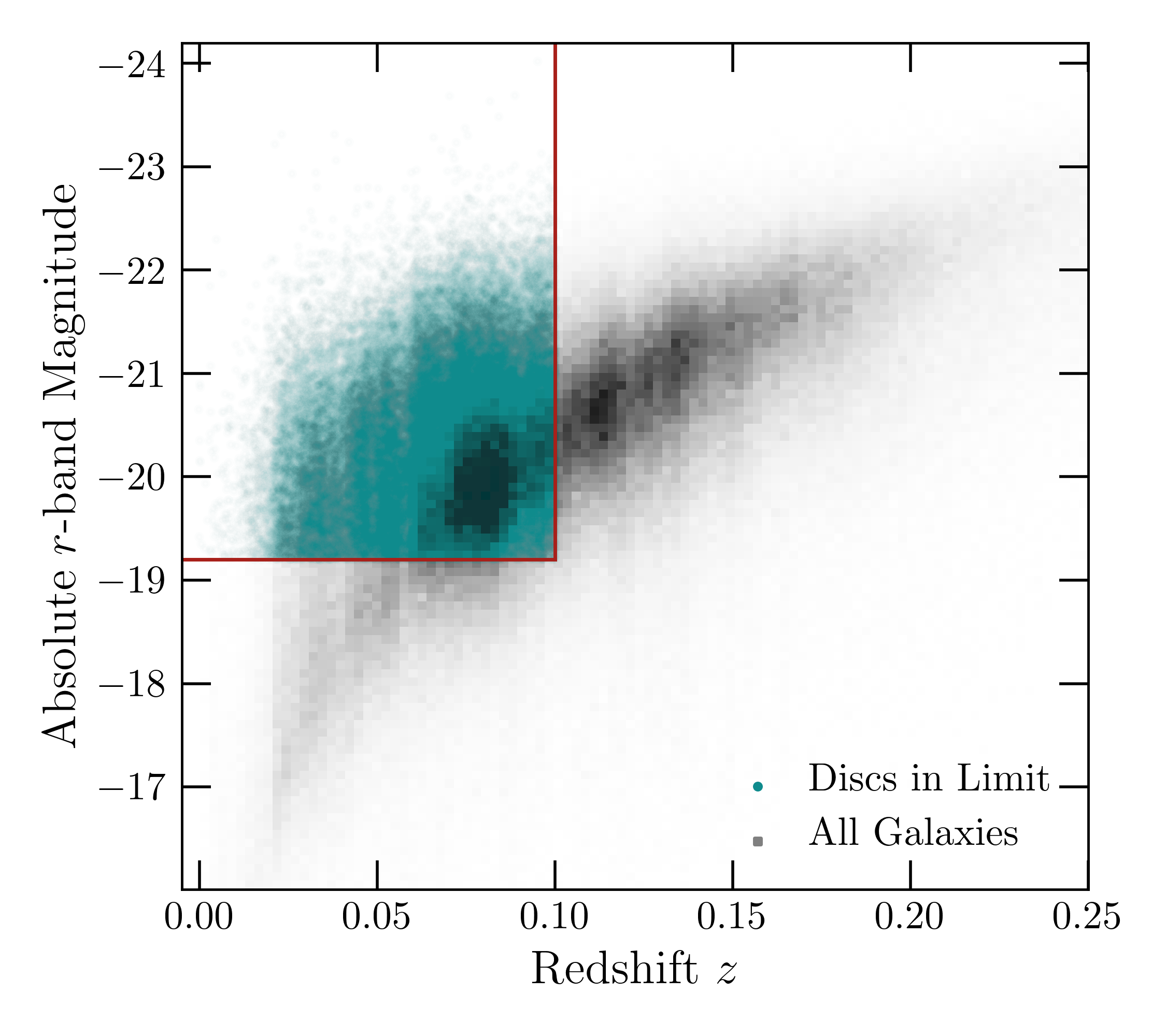}
    \caption{Absolute $r$-band magnitude against redshift, showing our volume limited sample. The grey-scale 2D histogram represents all galaxies in GZ DESI, and the teal points represent disc-dominated, not edge-on, merger-free galaxies within our volume limit. These teal points make up our full sample. The red lines at $M_{r} = -19.2$ and $z=0.10$ delineate our redshift and $r$-band magnitude limits.}
    \label{fig:mag_z_vollim}
\end{figure}

\begin{figure*}
    \centering
	\includegraphics[width=\textwidth]{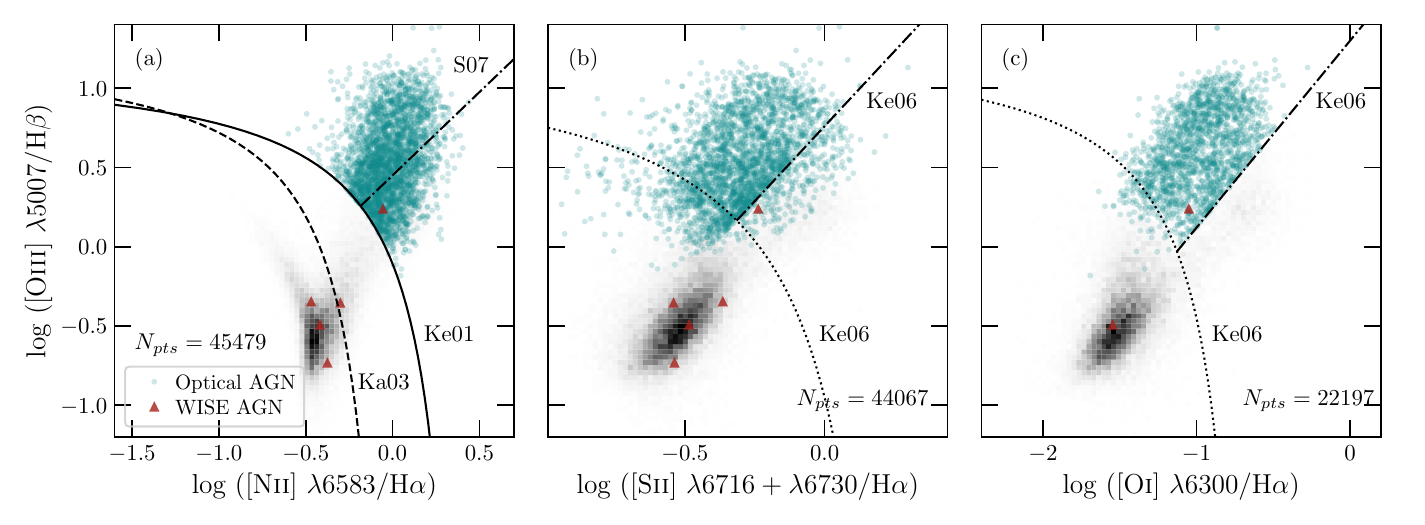}
    \caption{Classification of disc-dominated sources on a trio of emission line ratio diagrams \citep{baldwin1981, veilleux1987}. The grey histogram represents anything classed `star-forming', `LINER' or `composite', the teal points represent optically classified AGN and the red triangles represent WISE-classified AGN. From panel (a), we assume that any source falling below the Ka03 line \citep{kauffmann2003} is purely star-forming. Anything above the Ke01 line \citep{kewley2001} is either an AGN or a LINER, and thus any source lying between those two lines is classed as composite. To distinguish between AGNs and LINERs, we use Panels (c), then (b), then (a) in that order. This is because where a source has S/N > 3 in \oi, then Panel (c) is the most reliable, and we consider any source lying above the Ke06 \citep{kewley2006} line to be an AGN, whereas a source below is a LINER. Where a source has S/N < 3 in \oi, but S/N > 3 in \sii, we use Panel (b). Again, a source lying above the Ke06 line is classified as an AGN and below is a LINER. Where both \sii and \oi in a source have S/N < 3, we use \nii. Any source lying above the S07 line \citep{schawinski2007} is classified an AGN, and below is a LINER.}
    \label{fig:bpt_diagrams}
\end{figure*}

\subsection{Morphology Classification}
In order to examine the secular growth, we select galaxies which have a substantial disc component using GZD.
The first classification the model must perform is to select whether the galaxy is ``smooth and featureless'', has ``features or a disc'', or contains (or is) an ``artefact''.
To select disc galaxies, we require that the vote fraction for ``features or disc'' is $f_{\mathrm{smooth-or-featured\_featured-or-disk}} \geq 0.27$.

We also require that any discs must not be edge-on so that a bar can be identified if present, since in an edge-on galaxy, the bar can be obscured.
GZD must categorise each featured galaxy as ``edge-on'' or ``not edge-on'', and for our purposes, we require $f_{\mathrm{disk-edge-on\_no}} \geq 0.68$.
\citet{galloway2015} examine the relationship between inclination angle and observed bar fraction, and show (their Fig. 2) that the exact threshold used for ``not edge-on'' does not have a significant effect.
These limits follow those used in \citet{geron2021} and \citet{walmsley2022}.

To complete our sample, we require that the galaxy in the image does not appear to be merging with another galaxy.
GZD classifies every image with a merger class of ``merger'', ``major disturbance'', ``minor disturbance'', or ``none''.
We consider galaxies with any significant level of disturbance to be potential contaminants to a sample of discs undergoing secular evolution.
Thus we create a parameter we refer to as merger prominence, $\zeta_{\mathrm{avg}}$, analogous to the bulge prominence parameter in \citet{masters2019}.
We define $\zeta_{\mathrm{avg}}$ as:
\begin{align}
\nonumber    \zeta_{\mathrm{avg}} = 0.2 \times &f_{\mathrm{merging\_minor-disturbance}} \\
                               + 0.8 \times &f_{\mathrm{merging\_major-disturbance}} \\
\nonumber                      + &f_{\mathrm{merging\_merger}}
\end{align}
We require our sample to contain only galaxies which are not merging, which we identify as $\zeta_{\mathrm{avg}} < 0.3$.
This value has been visually checked to be consistent with undisturbed galaxies, via spot checking of $\sim50$ galaxies around the selected cut-off value.

In order to reduce selection effects, we select a volume-limited sample having $z \leq 0.10$ and and $M_{r} \leq -19.2 $, as shown in Fig. \ref{fig:mag_z_vollim}.
The 48,871 galaxies that form our final, complete sample (i.e., within the volume limit, disc-dominated, not edge-on, not merging) are shown in teal.

Within this volume-limited sample, we subsequently identify whether each of our galaxies has a bar, and the strength of that bar.
GZD asks the models to distinguish between ``strongly barred'', ``weakly barred'', and ``not barred''.
We classify a galaxy as unbarred if $f_{\mathrm{strong-bar+weak-bar}} < 0.5$.
We then divide the barred galaxies into strong and weak bars in order to investigate the effect of bar strength on AGN presence.
We define a barred galaxy as strongly barred if $f_{\mathrm{strong-bar}} \geq f_{\mathrm{weak-bar}}$, and weakly barred if $f_{\mathrm{strong-bar}} < f_{\mathrm{weak-bar}}$.
These limits follow the criteria used successfully in \citet{geron2021, geron2023}.
This means that every galaxy in our volume-limited disc sample is categorised as unbarred (\ubar, 27,391 galaxies), strongly barred (\sbar, 7,069 galaxies) or weakly barred (\wbar, 14,411 galaxies).}

As with any measurement, the GZD vote fractions do have errors associated with them. When the vote fractions are varied within their errors (assumed to be Gaussian) using a bootstrapping method iterated 1000 times with replacements, our results do not change.

\subsection{Activity Classification}

We use emission line ratio diagrams \citep*{baldwin1981, veilleux1987} to classify the galaxies in our sample as either: undetermined, star-forming, composite, AGN, or LINER.
We use the emission lines from MPA-JHU DR7 to place galaxies on the diagram, and we show the distribution in Fig. \ref{fig:bpt_diagrams}.

In order for a source to be classifiable according to this method, we require that the signal-to-noise ratio (S/N) in \oiii, \hb, \nii\ and \ha\ be $\mathrm{S/N} \geq 3$, in order to ensure good quality emission lines.
If a galaxy does not fulfil this first requirement, it may still be classifiable depending on where the limits lie \citep[e.g.,][]{brinchmann2004, salim2007, rosario2016}, which we discuss below.
If a source fulfils this requirement only in \ha\, it is classified as undetermined.

If a galaxy does fulfil all the S/N requirements, we use Panel (a) in Fig. \ref{fig:bpt_diagrams}, to classify a galaxy as star-forming if it falls below the Ka03 line \citep{kauffmann2003}.
If a galaxy falls between the Ka03 line and the Ke01 line \citep{kewley2001}, then it is classified as composite, and the emission is likely due to a combination of star formation and AGN/LINER activity.
If a galaxy falls above the Ke01 line, we classify it as either an AGN or a LINER (Low-Ionisation Nuclear Emission-line Region). We explain how these two objects are differentiated below.

If a source only fulfils the S/N criteria in \ha, \oiii\ and \hb, this provides an upper limit on the \ha/\nii\ ratio. Thus if this source falls below the Ka03 line, it is still classifiable as star-forming. Else, it is classified as uncertain.
If a source only fulfils the S/N criteria in \ha, \nii\ and \hb, this provides an upper limit on the \oiii/\hb\ ratio. Thus if this source falls below the Ka03 line, it is still classifiable as star-forming. Else, it is classified as uncertain.
If a source only fulfils the S/N criteria in \ha, \nii\ and \oiii, this provides a lower limit on the \oiii/\hb\ ratio. Thus if this source falls above the Ke01 line, it is still classifiable as either an AGN or a LINER. Else, it is classified as uncertain.
If a source only fulfils the S/N criteria in \ha\ and \hb, this provides a lower limit on the \nii/\ha\ ratio, and an upper limit on the \oiii/\hb\ ratio. Thus if this source below above the Ke01 line, it is still classifiable as starforming. Else, it is classified as uncertain.

There are three different emission lines we can use to distinguish AGNs from LINERs -- \sii, \oi\ and \nii.
The most reliable line is \oi\ \citep{kewley2006} and this should be used where possible, so if $\mathrm{S/N}_{\moi} \geq 3$, we can use Panel (c), and classify any source falling below the Ke06 line \citep{kewley2006} as a LINER.
This results in the hard cut-off line we see in Panel (c) that is not present in (a) or (b) for distinguishing between AGNs and LINERs.
Where $\mathrm{S/N}_{\moi}$ is too low, \sii\ is the next best emission line, and so if $\mathrm{S/N}_{\msii} \geq 3$ we can use Panel (b), and classify any source falling below the Ke06 line as a LINER.
Where both $\mathrm{S/N}_{\msii}$ and $\mathrm{S/N}_{\moi}$ are too low, we can resort to Panel (a), and use the S07 line \citep{schawinski2007}, since a source must have $\mathrm{S/N}_{\mnii} \geq 3$ in order to be  classified as either an AGN or a LINER at all.
Anything both below this line and above the Ke01 line can be classified as an LINER.

This leaves our volume-limited disc sample with: 712 undetermined galaxies, 2,518 uncertain galaxies, 28,807 starforming galaxies, 8,669 composite galaxies, 4,843 LINERs and 3,160 optically classified AGNs. When the fluxes are varied within their errors (assumed to be Gaussian) using a bootstrapping method iterated 1000 times with replacements, our results do not change.

Some AGNs are not optically classifiable, and are instead observable primarily in the infrared regime.
We match our catalogue to the Wide-Field Infrared Survey Explorer (WISE) AGN catalogue \citep{assef2018}, and any AGNs which are present in this catalogue, but not classified as AGNs according to the method described above, we add to our sample.
There are 5 WISE AGNs which appear in our volume-limited galaxy sample, 1 of which is classified as an AGN using emission line ratio diagrams, so we can reclassify an additional 4 galaxies as AGN, to give us a total of 3,164 AGNs.

\begin{figure}
    \centering
	\includegraphics[width=\columnwidth]{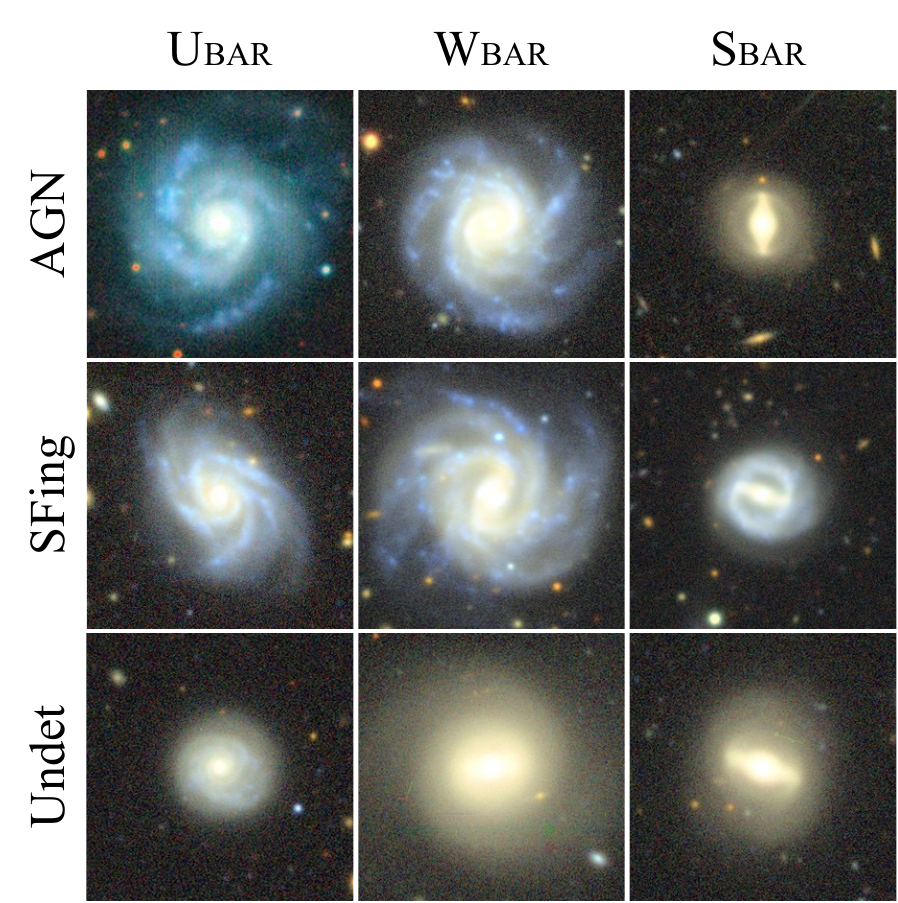}
    \caption{Examples of each morphology and activity classification. The left-hand column shows unbarred galaxies, the middle shows weakly barred, and the right-hand shows strongly barred galaxies.The top row shows AGN-host galaxies, the middle row shows star-forming galaxies, and the bottom row shows undetermined galaxies according to classification using emission line ratio diagrams. The undetermined galaxies are predominantly red spirals.}
    \label{fig:morphology_egs}
\end{figure}

Examples of different bar strengths in star-forming, AGN-host, and undetermined galaxies are shown in Fig. \ref{fig:morphology_egs}.
For a complete breakdown of how many galaxies are in each morphology category, and in each activity category, see Table \ref{tab:subsets_vollim} in Appendix \ref{sec:full_table}.
Note that whilst the classification of LINERs, uncertain sources and composite galaxies is important, it is simply so they can be confidently removed from our sample for analysis. We do not explicitly make use of these galaxies, as they are possible contaminants in our otherwise pure sample of Seyferts.

\section{Results}\label{sec:results}

\begin{figure}
    \centering
	\includegraphics[width=\columnwidth]{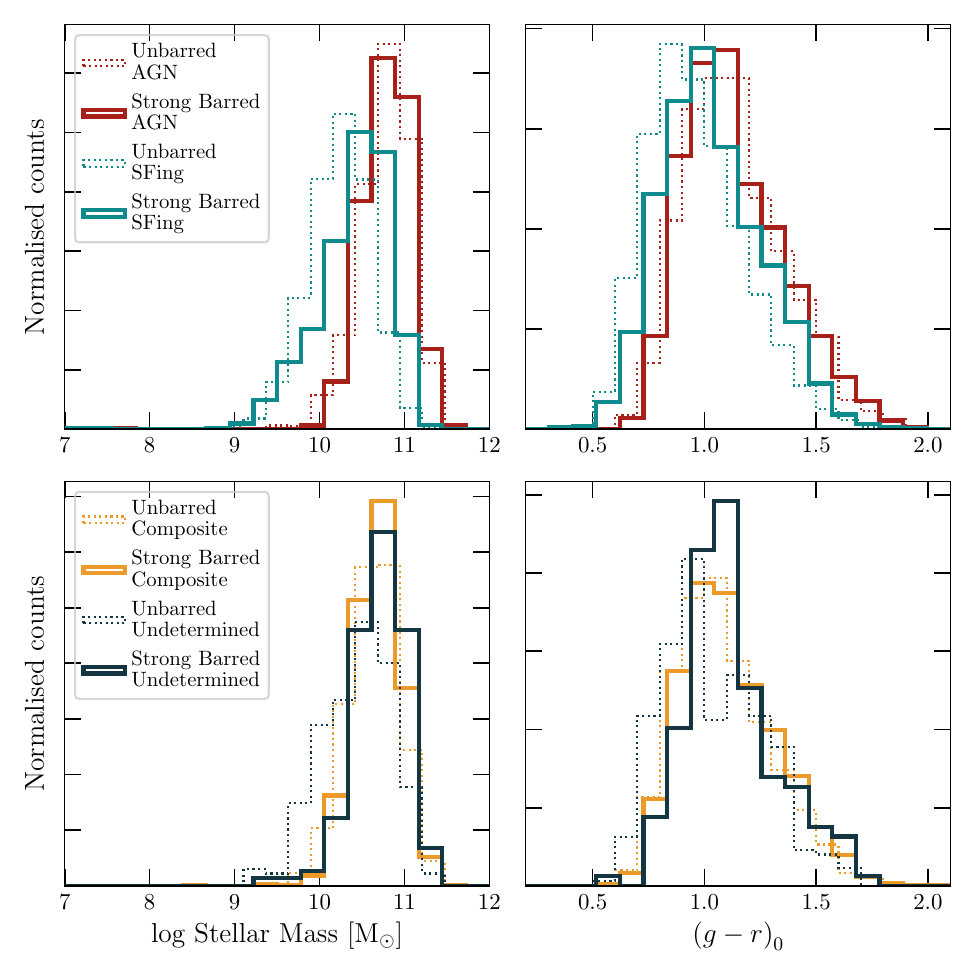}
    \caption{The distributions in \mstar\ and $(g-r)_{0}$ colour for a variety of subsamples, with strongly barred galaxies in solid lines, unbarred in dashed lines, AGNs in red, star-forming in teal, composite in orange and undetermined in navy blue. Weakly barred galaxies are not shown for simplification, but lie between the strongly barred and unbarred samples.}
    \label{fig:agn_fix_mass_col}
\end{figure}

We look at the variation in stellar mass (\mstar) and $(g-r)_{0}$ colour, where the 0 indicates we have corrected the $(g-r)$ colour for galactic absorption, between strongly barred (\sbar), weakly barred (\wbar) and unbarred galaxies (\ubar), and the results are shown in Fig. \ref{fig:agn_fix_mass_col}.
For visualisation purposes, we omit the results for \wbar\ since they lie between the two other samples (however this inclusive plot is shown in Fig. \ref{fig:agn_fix_mass_col_all} in Appendix \ref{sec:full_dist}).
As expected, the star-forming galaxies are less massive and slightly bluer than the AGN hosts.
The composite galaxies have overlap with both star-forming and AGN hosts, which confirms that their activity is due to a mixture of star formation and AGN.
This is very similar to the undetermined galaxies, whose signal to noise is too low to classify their activity.
The undetermined galaxies are predominantly a mix of quenching and fully quenched disc galaxies.
There are also some small differences between the barred and unbarred samples, with bars tending to reside in more massive, redder discs, particularly in both the star-forming samples and the undetermined samples, in agreement with previous studies \citep[e.g.,][]{masters2011}.

For further analysis, we limit our sample to only star-forming, undetermined and AGN host galaxies to avoid any ambiguity from the LINER and composite samples.

\begin{figure}
    \centering
	\includegraphics[width=\columnwidth]{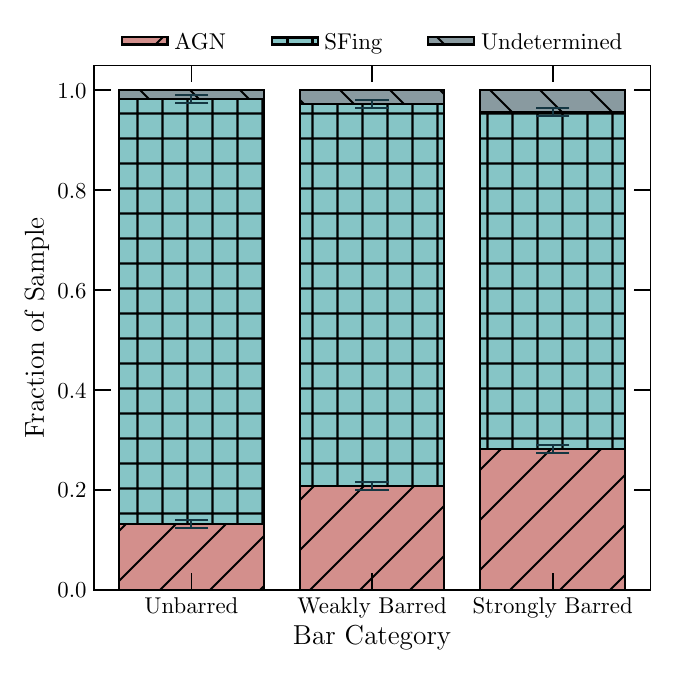}
    \caption{The distribution of activity classification within each bar category, as shown in Tab. \ref{tab:activity_distribution} AGN fraction is shown as positive diagonal in red, star-forming (SFing) is shown as teal square hatching, and Undetermined is shown as navy blue negative diagonal. Whilst in all three bar categories, the AGN fraction is smaller than the inactive fraction, the strongly barred galaxies have a noticeably greater fraction of AGN than the weakly barred galaxies, which in turn have a greater AGN fraction than the unbarred galaxies.}
    \label{fig:activity_distribution}
\end{figure}

We divide our sample of star-forming, undetermined and AGN host galaxies into our \sbar, \wbar\ and \ubar\ samples.
Within these three samples, we divide the \mstar\ and colour each into 15 bins of equal width (over $7.0 \leq \log(\mmstar/\mmsun) \leq 12.0$ and $0.4 \leq (g-r)_{0} \leq 2.0 $), and assign weights to each galaxy such that the weighted distributions of \mstar\ and colour are matched between the \sbar, \wbar\ and \ubar\ subsamples.
This is because AGN presence is known to correlate with \mstar\ and colour, and we want to reduce selection effects, and ensuring that the distributions are the same will aid this.
We can then determine the fraction in each bar category of AGN, star-forming and undetermined galaxies. However, when determining the activity fractions, we introduced a more conservative mass cut of $10.0 \leq \log(\mmstar/\mmsun) \leq 12.0$ before weighting the distributions. This is because AGN are more easily observable in higher mass galaxies \citep{aird2012}, and this conservative mass cut reduces this selection bias.
Changing this lower mass limit between $10^{9.0}\mmsun$ and $10^{10.0}\mmsun$ does not change our results.
The weighted results are shown in Fig. \ref{fig:activity_distribution}, and Table \ref{tab:activity_distribution}.
Errors arise from the binomial distribution \citep{cameron2011}.

\begin{table}
    \caption{The percentage of each activity category within each bar classification, as shown in Fig. \ref{fig:activity_distribution}. AGN presence in strongly barred galaxies is around twice as prolific as in weakly barred or unbarred galaxies.}
    \label{tab:activity_distribution}
    \begin{tabular}{rccc}\hline
        \multicolumn{1}{l|}{} & Strongly Barred & Weakly Barred & Unbarred      \\ \hline
        AGN                   & $31.6\pm0.9$    & $23.3\pm0.8$  & $14.2\pm0.6$  \\
        Star-forming          & $63.6\pm0.9$    & $73.6\pm0.8$  & $83.9\pm0.6$  \\
        Undetermined          & $ 4.7\pm0.4$    & $ 3.1\pm0.3$  & $ 1.9\pm0.2$  \\ \hline
    \end{tabular}
\end{table}

Given the small errors on each of these fractions within each bar category, it is highly unlikely that any of these subsamples are drawn from the same parent distribution.
These initial results indicate that AGNs are more likely to reside in strongly barred galaxies than in weakly barred, and even less likely to reside in unbarred.
However, given the ranges of \mstar\ and colour, we endeavour to examine these fractions as a function of both, whilst simultaneously examining how the AGN fractions may vary across \mstar-colour space.
Given that we cannot classify the star-formation rate of the AGN hosts, we do not consider SFR as a parameter, and thus we combine star-forming galaxies and undetermined galaxies into one category, which we refer to as `inactive' galaxies.

\subsection{AGN-bar correlation with stellar mass and colour}\label{sec:agn_bar_as_mass_colour} 
\begin{figure*}
    \centering
	\includegraphics[width=\textwidth]{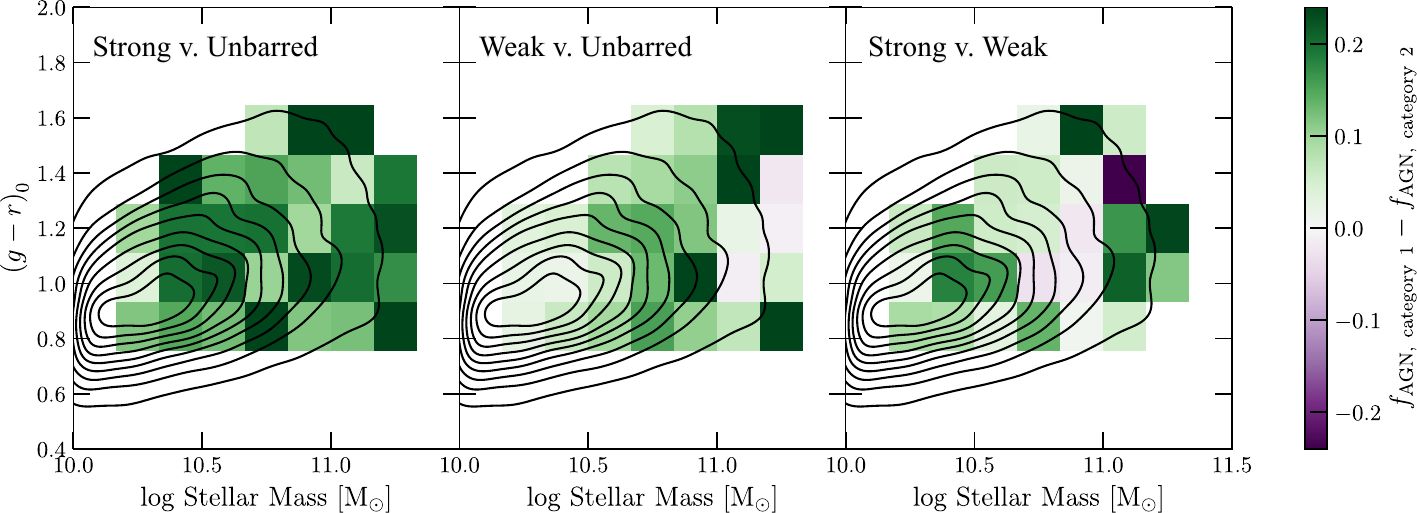}
    \caption{The difference between the AGN fraction in two bar categories for every combination of \sbar, \wbar\ and \ubar\, with \mstar\ on the x-axis and $(g-r)_{0}$ colour on the y-axis. In each case, the label along the top is written as `Category 1 v. Category 2'. The black contours indicate the population of disc galaxies (AGN-host and inactive) within the volume limit. The 2D histogram indicates the distribution of AGN-host disc galaxies, where there are a minimum of 17 AGN in a bin. Where the bin is more green, this indicates that the fraction of Category 1 galaxies hosting AGNs is greater than the fraction of Category 2 galaxies hosting AGNs. Where the bin is more purple, the reverse is true.}
    \label{fig:fagn-fnoagn}
\end{figure*}

We divide our sample of AGN hosts and inactive galaxies into nine bins in \mstar\ and nine bins in $(g-r)_{0}$ colour.
Within each bin, we calculate the AGN fraction in strongly barred galaxies, $\mfagnsbar$, and the AGN fraction in unbarred galaxies, $\mfagnubar$.
We then find the difference in these two fractions, and this is shown in Fig. \ref{fig:fagn-fnoagn}a. It can be assumed that there is one, real, intrinsic value for this difference, but when we sample it, we get some scatter. Thus, by sampling it multiple times by varying the binning, we should get an approximately Normal distribution that centres around the true value.

The difference between the two fractions is shown as a colour bar, where green indicates that the fraction of strongly barred galaxies which host AGNs is greater than the fraction of unbarred galaxies which host AGNs.
In order to reduce noise, we only show bins where there are at least 17 AGNs in a bin.
Varying the minimum AGN count per bin within reasonable values does not change our qualitative result.
Every bin is green, with approximately $\mfagnsbar - \mfagnubar \approx 0.18$.
This is a small but significant increase in the number of AGNs in strongly barred galaxies.

When we repeat this analysis for \wbar\ and \ubar\, our result is qualitatively similar, in that weakly barred galaxies appear more likely to host an AGN.
But the signal is much less strong, indicating that any effect that weak bars have on AGN presence is less pronounced than for strong bars, which is also reflected in the lower fractions over the full \mstar-colour-matched sample.
This is plotted in Fig. \ref{fig:fagn-fnoagn}b.
Overall the AGN fraction is still higher in weakly barred galaxies than in disc galaxies without bars.

We can directly compare \sbar\ and \wbar\ across the \mstar-colour diagram as well (Fig. \ref{fig:fagn-fnoagn}c), and we find that the fraction of strongly barred galaxies hosting AGNs is significantly greater than the fraction of weakly barred galaxies hosting AGNs although again this is less pronounced than in Fig. \ref{fig:fagn-fnoagn}a.

\begin{figure}
    \centering
	\includegraphics[width=\columnwidth]{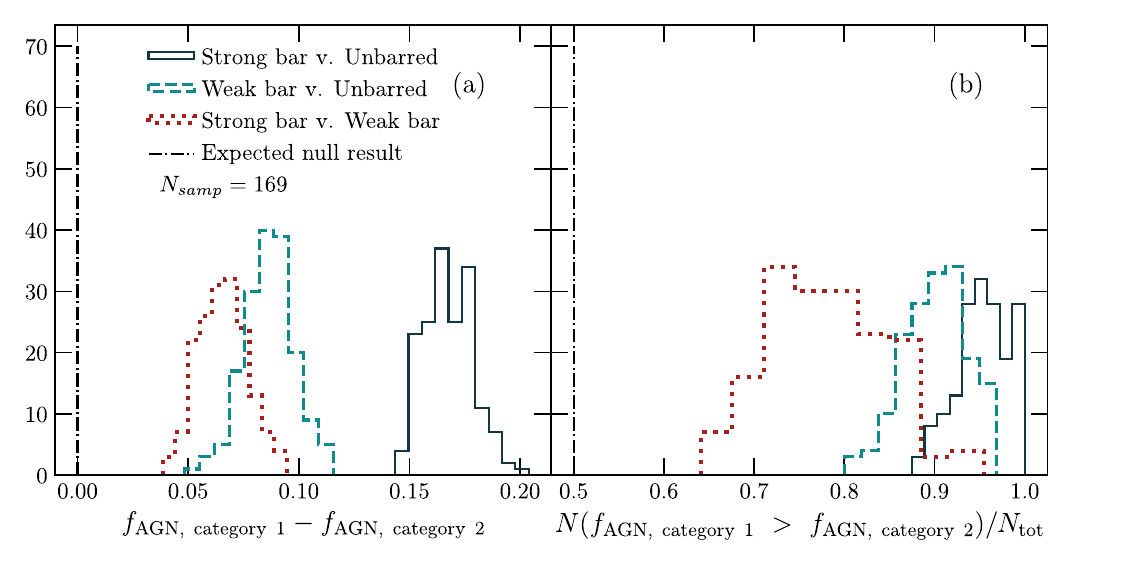}
    \caption{The distributions of the median difference between AGN fractions in Panel (a), and the distribution of the fraction of bins where Category 1 bars are greater than Category 2, in a sample consisting of AGN-host galaxies and inactive galaxies. In each case, the legend is written as `Category 1 v. Category 2'. The black dashed-dotted lines indicate the expected mean of the distributions if bar presence did not affect AGN presence. The navy blue, solid lines represent \sbar\ v. \ubar. The teal, dashed lines represent \wbar\ v. \ubar. The red, dotted lines represent \sbar\ v. \wbar. The further to the left of the expected null result the histograms lie, the greater the tendency for AGN to lie in bar Category 1 galaxies.}
    \label{fig:df_hist}
\end{figure}

Given that any increase in AGN fraction is small, we check that this value is not overly dependent on binning, and we repeat these calculations for every \mstar\ and colour bin combination from 5 bins to 17 bins, for a total of 169 bin combinations.
For each binning combination, we calculate the median difference in AGN fraction (e.g., $\mfagnsbar - \mfagnubar$), and we plot these medians in Fig. \ref{fig:df_hist}a.
We assume that the different binning choices each sample the true value of the difference in AGN fraction between subsamples, such that the distribution of values recovered from all binning choices represents the measured value and its uncertainty. This means that the histogram for each comparison does not consist of independent measurements, but rather is expected to peak around the true value of the difference in the AGN fraction.

If there was no difference in the likelihood of hosting an AGN between these three subsamples, we would expect the histograms to centre around 0 (e.g., $\mfagnsbar - \mfagnubar = 0$).
We always take the weaker bar category from the stronger bar category, so if the centre of the histograms is greater than 0, the stronger bar category is more likely to host an AGN than the weaker, and vice versa if the centre of the histogram is less than 0.

Fig. \ref{fig:df_hist}a shows that the stronger bar category is more likely to host an AGN than the weaker bar category in every case: strongly barred galaxies are more likely to host an AGN than weakly barred galaxies, which are in turn more likely to host an AGN than unbarred galaxies.
Yet this excess of AGNs we see is very small.
The difference in AGN fraction between strongly barred and unbarred galaxies is $0.17\pm0.01$, between strongly barred and weakly barred is $0.07\pm0.01$, and between weakly barred and unbarred is $0.09\pm0.01$.

A Shapiro-Wilk test \citep{shapiro1965} shows that we cannot reject Normality for any of the distribution of the medians, with p-values in each case greater than $p_{\mathrm{SW}}>0.21$ ($<1.3\upsigma$).
Given these distributions are consistent with the Normal distribution, we can perform a simple T-test \citep{student1908} to quantify the significance of this excess of AGN.
In each of these cases, the p-value resulting from a T-test is $p_{\mathrm{T}} \ll 1\times10^{-6}$ ($ \gg 5\upsigma$), and thus we reject the hypothesis that the likelihood of each of these bar categories hosting an AGN are identical to each other.
Furthermore, we can say that the galaxies in \ubar\ are less likely to host an AGN than the galaxies in \sbar\ or \wbar\ category to a $5\upsigma$ confidence.

For each binning combination, we also calculate the fraction of bins where the stronger bar category hosts a greater AGN fraction than the weaker bar category (e.g.$\mfagnsbar > \mfagnubar$), and we plot these values in Fig. \ref{fig:df_hist}b.

For example, if we have 5x5 bins, and 20 bins have $\mfagnsbar > \mfagnubar$, we would report a value of 0.8 for this bin combination.
If there was no difference in the likelihood of hosting an AGN between our three subsamples, we would expect the distributions to centre around 0.5 -- half of the bins would show a greater fraction of AGNs in one bar category than the other.
This point is signified by a dash-dotted line.

Comparing \sbar\ and \ubar, the fraction of bins where \sbar\ has a greater AGN fraction is $0.95\pm 0.03$.
For \sbar\ and \wbar\ this is $0.78\pm 0.06$, and for \wbar\ versus \ubar, the fraction of bins where \wbar\ has a greater AGN fraction is $0.90\pm 0.04$, where errors arise from the standard deviation.

Again, we perform a Shapiro-Wilk test for Normality.
For the combination \wbar\ vs. \ubar, we obtain a p-value of $p_{\mathrm{SW}}=0.10$ ($1.6\upsigma$), and thus for this comparison, we can use a T-test to quantify the significance of the excess of bins containing a higher AGN fraction in the weakly barred category.
The p-value resulting from this T-test is $p_{\mathrm{T}} \ll 1\times10^{-6}$ ($\gg 5\upsigma$).

Since we can reject Normality for \sbar\ vs. \ubar, and \sbar\ vs. \wbar, ($p_{\mathrm{SW}} \leq 0.03$), we must use the more conservative method of calculating the number of standard deviations between the mean and the null result of 0.5. 
For \sbar\ v. \ubar, with a mean value of 0.95, and a standard deviation, $\sigma_{\mathrm{SD}}$, of 0.03, we can say that the mean is $15\sigma_{\mathrm{SD}}$ away from 0.5, and therefore is not in agreement.
For \sbar\ v. \wbar, with a mean value of 0.78, and a standard deviation, $\sigma_{\mathrm{SD}}$, of 0.06, we can say that the mean is $13\sigma_{\mathrm{SD}}$ away from 0.5, and therefore is not in agreement.
Thus, in each case, the stronger bar category has an AGN fraction that is greater than the weaker bar category.
This occurs across the \mstar-colour regime, meaning that there is not one specific combination of \mstar\ and colour driving this relationship, further justifying that our results are not sensitive to the choice of binning.

The trends between bar strength and AGN activity are likely a mix of relatively straightforward and more complex results. We discuss these further below.

\section{Discussion}\label{sec:discussion}
Our overall result, with AGN activity in both unbarred and barred disc galaxies, confirms that a large-scale bar is not \emph{required} to feed an AGN in the secular-evolution regime.
There are multiple secular channels by which matter from the kiloparsec-scale disc can flow into the SMBH sphere of influence, according to both simulations \citep[e.g.,][]{ciotti1991,friedli1993,sakamoto1996,maciejewski2002,regan2004,hopkins2006b,ciotti2012,lin2013, slater2019} and observations \citep[e.g.,][]{davies2007, smethurst2021}. 
However, our primary result also shows clear evidence for an increase in AGN activity in both strongly and weakly barred systems, and we focus on discussing this result below.

As described in Section \ref{sec:results}, we find that strong bars are clearly linked to a higher incidence of AGN activity, and that weak bars show a more subtle, but still positive, correlation. These results clarify the debate over the last few years regarding whether (and how much) bars are associated with AGN activity.
They also highlight an emerging consensus regarding the link between bars and AGN. For example, our results agree with \citet{silva-lima2022}, who counter for selection effects and find that barred galaxies have a higher accretion parameter than unbarred, and that AGNs are found more commonly in galaxies with a bar.
Since we are looking at incidence rather than luminosity or accretion, this study is particularly complementary.
Given that recent studies have shown that strong and weak bars must be considered separately \citep{geron2023}, this could also be responsible for some of the discrepancies seen in contradicting previous studies \citep[e.g.,][]{cheung2013, goulding2017, Zee2023}, who find no correlation between bars and AGNs.

It is natural at this point to consider inverting the question we have been considering throughout this work, and instead examine the fraction of AGN that have bars. Within our sample and the defined mass limits, $40.0\pm1.0$ per cent of AGN have strong bars, $34.7\pm1.0$ per cent have weak bars, and the remaining $25.3\pm1.0$ per cent are unbarred. $24.5\pm0.5$ per cent of inactive galaxies have strong bars, $43.2\pm0.6$ per cent have weak bars, and the remaining $32.3\pm0.5$ per cent are unbarred. This agrees qualitatively with multiple previous studies, such as \citet{oh2012} and \citet{garland2023}, although the exact numbers can vary with the study because of the sample selection and bar sensitivities of each individual study. This is the largest study so far to examine the disc galaxy population for evidence of correlation between bars and AGN and there is clearly a positive correlation in both directions. AGN are most likely to host strong bars, and strongly barred galaxies are most likely to host AGN, within the disc galaxy population.

Our findings are consistent with recent evidence that strong and weak bars have different formation mechanisms \citep[e.g.,][]{geron2023}: strong bars are triggered by global disc instabilities, whereas weak bars are formed through tidal interactions.
These formation mechanisms could both be responsible for triggering an AGN. Thus, the AGN's presence may not be directly due to the bar, but rather to the same mechanisms that caused the bar to form.
If the physical mechanisms are different for strong and weak bars, this leads to a different co-incidence between AGN and the two different bar strengths.
Tidal interactions may be more efficient at depleting gas from the centre of the galaxy than secular processes that do not lead to the development of a bar, such that by the time the disc galaxy has evolved to a higher stellar mass, the AGN has been deprived of fuel and shut down, although the weak bar still remains in place.

The gas content is significantly higher in starforming discs than red spirals \citep{masters2012}.
We postulate that were a strong bar present in a galaxy, this bar is efficient at fuelling this additional gas down to the central kilo-parsec, where it can be accreted onto an AGN.
If weak bars are less efficient at driving gas to the centre of the galaxy, this would explain why we see a much weaker correlation between weakly barred galaxies and AGNs than strongly barred galaxies and AGN.

There is recent evidence from IllustrisTNG that an AGN could drive changes in the bars \citep{lokas2022}. This is due to the AGN switching to kinetic-mode feedback, causing depletion of gas in the inner regions, leading to quenching and bar-formation. However given the physical scales that we are looking at, this seems unlikely to be occurring on a large-scale in our sample, due to the differences in AGN and bar lifetimes.

It is important to assess the potential contribution to this result of any selection biases. As described in Section \ref{sec:data_collection}, we take various steps to minimise these biases, such as controlling for \mstar\ and $g-r$ colour, and using a volume limit to ensure completeness. We also consider the effects of the changing physical resolution across the sample. At higher redshifts, the minimum size of bar we can detect increases: we lose smaller bars at higher redshift.
Weak bars tend to be shorter than strong bars proportional to the size of the galaxy.
This means we preferentially lose weaker bars as we increase in redshift, especially in lower mass galaxies.
In a low mass galaxy, a weak bar may be missed, and that galaxy classified as unbarred.

We do not have individual bar lengths and widths for all the galaxies in this sample, and thus we cannot fully compensate for this potential source of selection bias at an individual galaxy level.
However, one way to examine how strong this effect is likely to be in our sample is to remove the high redshift sources, because this will significantly reduce the overall difference between minimum resolved bar size between the lowest and highest redshifts of the subsample.
We have thus examined the subset of our volume-limited sample with $z < 0.05$ (12,251 disc galaxies), and the overall trends seen in Fig. \ref{fig:fagn-fnoagn} still persist.

Further work could be done to investigate the inflow rates that each of these bar types could sustain, and combining this with the gas availability could show why weak bars do not correlate with AGN presence as much as strong bars, as if they cannot provide as high an inflow rate as strong bars, they require more gas to trigger an AGN.
High-resolution IFU data will allow us to measure star-formation rates of the AGN-host galaxies, and thus draw comparisons between AGN and non-AGN hosts, both in starforming and quiescent galaxies.
X-ray data will allow investigation of black hole accretion rates, and spectroscopy along the axes of the bar will allow bar inflow rates to be obtained.

This phenomenon will also be investigated at more distant redshifts \citep[][Margalef-Bentabol et al., in prep.]{margalefbentabol2023}, along with how these AGNs are fuelled as the bar fraction decreases out to higher redshifts.
Facilities such as Euclid will provide us with greater sky coverage at better resolution than currently available, and so with an increase in data, we should be able to reduce noise in our samples.

\section{Conclusions}\label{sec:conclusions}
We have investigated the influence of large-scale bars on the likelihood of AGN signals in a volume-limited sample of 48,871 disc galaxies by analysing data from the DESI catalogue, Galaxy Zoo DESI morphologies, and SDSS emission line strengths.
We have taken care to control for differences in stellar mass and galaxy colour distributions between subsamples of strongly barred, weakly barred, and unbarred galaxies.

99.9 per cent of our 3164 AGN in disc galaxies are identified via optical emission line diagnostics, with a mere 4 AGN only detectable via WISE infrared colours within our volume limit.
We divide galaxies without clear AGN activity into multiple categories based on the detection of emission lines in SDSS spectra, and focus our comparison with the AGN host galaxies on two inactive categories: 28,807 star forming galaxies with detected nebular emission lines below the ``composite'' limit on an emission line ratio diagram, and 712 ``undetermined'' galaxies where nebular emission lines are not robustly detected in the central fibre spectra.
These latter galaxies are, on visual inspection, predominantly red spirals, with a smaller fraction being discs that have red/quenched inner regions and bluer outer regions.

Our key findings can be summarised as follows:
\begin{itemize}
    \item Strongly barred galaxies are more likely to host an AGN than weakly barred galaxies, which in turn are more likely to host an AGN than unbarred galaxies.
    \item This effect is very slight, with the fraction of AGN in each bar category being: $\mfagnsbar = 31.6 \pm 0.9$, $\mfagnwbar = 23.3 \pm 0.0$ and $\mfagnubar = 14.2 \pm 0.6$.
\end{itemize}

The high levels of statistical significance achieved here even after controlling for the confounding effects of colour, stellar mass, and flux limits, have been facilitated by the advent of large sample sizes from the latest generation of extragalactic surveys and the highly accurate and detailed morphological identifications of strongly barred, weakly barred, and unbarred disc galaxies.
In the near future we expect to use data from surveys such as Euclid and LSST to extend these analyses to higher redshift and further refine our understanding of the interplay between various types of disc instabilities and growing supermassive black holes.

\section*{Data Availability}
The data from GZD is available in \citet{walmsley2023b}

Other catalogues used are publicly available from the following locations:
\begin{itemize}
    \item MPA-JHU: \\
    \url{https://wwwmpa.mpa-garching.mpg.de/SDSS/DR7/}
    \item NYU-VAGC:\\
    \url{http://sdss.physics.nyu.edu/vagc/}
    \item DESI Legacy Surveys:\\
    \url{https://www.legacysurvey.org/dr10/description/}
\end{itemize}

\section*{Acknowledgements}
We would like to thank the referee for their valuable comments and insight, which improved the quality of this paper

ILG acknowledges support from an STFC PhD studentship [grant number ST/T506205/1] and from the Faculty of Science and Technology at Lancaster University.
MW is a Dunlap Fellow. The Dunlap Institute is funded through an endowment established by the David Dunlap family and the University of Toronto.
MSS acknowledges support from an STFC PhD studentship [grant ST/V506709/1].
LMP acknowledges support from the Faculty of Science and Technology at Lancaster University.
BDS acknowledges support through a UK Research and Innovation Future Leaders Fellowship [grant number MR/T044136/1].
CJL acknowledges support from the Sloan Foundation. 
RJS acknowledges support through the Royal Astronomical Society Research Fellowship.
KBM acknowledges support from the National Science Foundation under grants OAC 1835530 and IIS 2006894.
DOR acknowledges the support of the UK Science and Technology Facilities
Council (STFC) under grant reference ST/T506205/1..
JP acknowledges funding from the Science and Technology Facilities Council (STFC) Grant Code ST/X508640/1.
MRT acknowledges support from an STFC PhD studentship [grant number ST/V506795/1].

Funding for SDSS-III has been provided by the Alfred P. Sloan Foundation, the Participating Institutions, the National Science Foundation, and the U.S. Department of Energy Office of Science.
The SDSS-III web site is \url{http://www.sdss3.org/}.

SDSS-III is managed by the Astrophysical Research Consortium for the Participating Institutions of the SDSS-III Collaboration including the University of Arizona, the Brazilian Participation Group, Brookhaven National Laboratory, Carnegie Mellon University, University of Florida, the French Participation Group, the German Participation Group, Harvard University, the Instituto de Astrofisica de Canarias, the Michigan State/Notre Dame/JINA Participation Group, Johns Hopkins University, Lawrence Berkeley National Laboratory, Max Planck Institute for Astrophysics, Max Planck Institute for Extraterrestrial Physics, New Mexico State University, New York University, Ohio State University, Pennsylvania State University, University of Portsmouth, Princeton University, the Spanish Participation Group, University of Tokyo, University of Utah, Vanderbilt University, University of Virginia, University of Washington, and Yale University.

The Legacy Surveys consist of three individual and complementary projects: the Dark Energy Camera Legacy Survey (DECaLS; Proposal ID \#2014B-0404; PIs: David Schlegel and Arjun Dey), the Beijing-Arizona Sky Survey (BASS; NOAO Prop. ID \#2015A-0801; PIs: Zhou Xu and Xiaohui Fan), and the Mayall z-band Legacy Survey (MzLS; Prop. ID \#2016A-0453; PI: Arjun Dey).
DECaLS, BASS and MzLS together include data obtained, respectively, at the Blanco telescope, Cerro Tololo Inter-American Observatory, NSF’s NOIRLab; the Bok telescope, Steward Observatory, University of Arizona; and the Mayall telescope, Kitt Peak National Observatory, NOIRLab.
Pipeline processing and analyses of the data were supported by NOIRLab and the Lawrence Berkeley National Laboratory (LBNL).
The Legacy Surveys project is honored to be permitted to conduct astronomical research on Iolkam Du’ag (Kitt Peak), a mountain with particular significance to the Tohono O’odham Nation.

NOIRLab is operated by the Association of Universities for Research in Astronomy (AURA) under a cooperative agreement with the National Science Foundation.
LBNL is managed by the Regents of the University of California under contract to the U.S. Department of Energy.

This project used data obtained with the Dark Energy Camera (DECam), which was constructed by the Dark Energy Survey (DES) collaboration.
Funding for the DES Projects has been provided by the U.S. Department of Energy, the U.S. National Science Foundation, the Ministry of Science and Education of Spain, the Science and Technology Facilities Council of the United Kingdom, the Higher Education Funding Council for England, the National Center for Supercomputing Applications at the University of Illinois at Urbana-Champaign, the Kavli Institute of Cosmological Physics at the University of Chicago, Center for Cosmology and Astro-Particle Physics at the Ohio State University, the Mitchell Institute for Fundamental Physics and Astronomy at Texas A\&M University, Financiadora de Estudos e Projetos, Fundacao Carlos Chagas Filho de Amparo, Financiadora de Estudos e Projetos, Fundacao Carlos Chagas Filho de Amparo a Pesquisa do Estado do Rio de Janeiro, Conselho Nacional de Desenvolvimento Cientifico e Tecnologico and the Ministerio da Ciencia, Tecnologia e Inovacao, the Deutsche Forschungsgemeinschaft and the Collaborating Institutions in the Dark Energy Survey.
The Collaborating Institutions are Argonne National Laboratory, the University of California at Santa Cruz, the University of Cambridge, Centro de Investigaciones Energeticas, Medioambientales y Tecnologicas-Madrid, the University of Chicago, University College London, the DES-Brazil Consortium, the University of Edinburgh, the Eidgenossische Technische Hochschule (ETH) Zurich, Fermi National Accelerator Laboratory, the University of Illinois at Urbana-Champaign, the Institut de Ciencies de l’Espai (IEEC/CSIC), the Institut de Fisica d’Altes Energies, Lawrence Berkeley National Laboratory, the Ludwig Maximilians Universitat Munchen and the associated Excellence Cluster Universe, the University of Michigan, NSF’s NOIRLab, the University of Nottingham, the Ohio State University, the University of Pennsylvania, the University of Portsmouth, SLAC National Accelerator Laboratory, Stanford University, the University of Sussex, and Texas A\&M University.

BASS is a key project of the Telescope Access Program (TAP), which has been funded by the National Astronomical Observatories of China, the Chinese Academy of Sciences (the Strategic Priority Research Program “The Emergence of Cosmological Structures” Grant \# XDB09000000), and the Special Fund for Astronomy from the Ministry of Finance.
The BASS is also supported by the External Cooperation Program of Chinese Academy of Sciences (Grant \# 114A11KYSB20160057), and Chinese National Natural Science Foundation (Grant \# 12120101003, \# 11433005).

The Legacy Survey team makes use of data products from the Near-Earth Object Wide-field Infrared Survey Explorer (NEOWISE), which is a project of the Jet Propulsion Laboratory/California Institute of Technology. NEOWISE is funded by the National Aeronautics and Space Administration.

The Legacy Surveys imaging of the DESI footprint is supported by the Director, Office of Science, Office of High Energy Physics of the U.S. Department of Energy under Contract No. DE-AC02-05CH1123, by the National Energy Research Scientific Computing Center, a DOE Office of Science User Facility under the same contract; and by the U.S. National Science Foundation, Division of Astronomical Sciences under Contract No. AST-0950945 to NOAO.

The data in this paper are the result of the efforts of the Galaxy Zoo volunteers, without whom none of this work would be possible.
Their efforts are individually acknowledged at \url{http://authors.galaxyzoo.org}.

\subsection*{Software}
This research has made use of \textsc{Topcat} \citep{taylor2005}, an interactive graphical tool for analysis and manipulation of tabular data.

This research has made extensive use of the following Python packages:
\begin{itemize}
\item \textsc{Astropy}, a community-developed core Python package for Astronomy \citep{robitaille2013, price-whelan2018, price-whelan2022}.
\item \textsc{Matplotlib}, a 2D graphics package for Python \citep{hunter2007}.
\item \textsc{Numpy} \citep{harris2020}, a package for scientific computing.
\item \textsc{Scipy} \citep{virtanen2020}, a package for fundamental algorithms in scientific computing.
\end{itemize}



\bibliographystyle{mnras}
\bibliography{bibliography.bib} 




\appendix

\section{Subsample Counts}\label{sec:full_table}
Tab. \ref{tab:subsets_vollim} presents the full set of number counts of all subsamples in this work. While our analysis is confined to the volume-limited sample, we also present numbers for the full set of GZD classified galaxies which also have ancillary data presented in the MPA-JHU and NYU-VAGC catalogues.

\begin{table}
    \centering
    \caption{Full breakdown of the number of galaxies in each activity class, and each bar category, both in the volume limited sample, and before volume limiting. Note that the numbers in some of the sub-sub-categories may have duplicates. For example, there is one WISE AGN that is also an optical AGN, and thus the numbers do not completely add, and we show the total numbers for clarity.}
    \label{tab:subsets_vollim}
        \begin{tabular}{lllcc}
        \hline
        \multicolumn{5}{c}{Subsample Counts}                                        \\ \hline\hline
                &           &                           & Total    & In Volume Limit\\
        \multicolumn{3}{l}{Is Disc}                     & 112699   & 48871          \\
        \multicolumn{3}{l}{Is Undetermined}             & 218101   & 25004          \\
        \multicolumn{3}{l}{Is Uncertain}                & 124990   & 29355          \\
        \multicolumn{3}{l}{Is Star-forming}             & 280867   & 86917          \\
        \multicolumn{3}{l}{Is Composite}                & 75872    & 30540          \\
                & \multicolumn{2}{l}{LINER \oi}         & 11724    & 5597           \\
                & \multicolumn{2}{l}{LINER \sii}        & 26699    & 13189          \\
                & \multicolumn{2}{l}{LINER \nii}        & 11298    & 4366           \\
        \multicolumn{3}{l}{Is LINER}                    & 49721    & 23152          \\
                & \multicolumn{2}{l}{Optical AGN}       & 37558    & 13394          \\
                & \multicolumn{2}{l}{WISE AGN}          & 98       & 14             \\
        \multicolumn{3}{l}{Is AGN}                      & 37651    & 13406          \\ \hline \hline
        \multicolumn{3}{l}{Is Disc and:}                &          &                \\
                & \multicolumn{2}{l}{Is Undetermined}   & 2652     & 712            \\
                & \multicolumn{2}{l}{Is Uncertain}      & 11011    & 2518           \\
                & \multicolumn{2}{l}{Is Star-forming}   & 66335    & 28807          \\
                & \multicolumn{2}{l}{Is Composite}      & 17754    & 8669           \\
                &           & LINER \oi                 & 2035     & 1420           \\
                &           & LINER \sii                & 4219     & 2835           \\
                &           & LINER \nii                & 1172     & 588            \\
                & \multicolumn{2}{l}{Is LINER}          & 7426     & 4843           \\
                &           & Optical AGN               & 6625     & 3160           \\
                &           & WISE AGN                  & 16       & 5              \\
                & \multicolumn{2}{l}{Is AGN}            & 6639     & 3164           \\ \hline
        \multicolumn{3}{l}{Is Unbarred Disc and:}       &          &                \\
                & \multicolumn{2}{l}{Is Undetermined}   & 1377     & 350            \\
                & \multicolumn{2}{l}{Is Uncertain}      & 5432     & 1234           \\
                & \multicolumn{2}{l}{Is Star-forming}   & 42045    & 18829          \\
                & \multicolumn{2}{l}{Is Composite}      & 7051     & 3628           \\
                &           & LINER \oi                 & 861      & 592            \\
                &           & LINER \sii                & 1724     & 1155           \\
                &           & LINER \nii                & 501      & 256            \\
                & \multicolumn{2}{l}{Is LINER}          & 3086     & 2003           \\
                &           & Optical AGN               & 2448     & 1237           \\
                &           & WISE AGN                  & 7        & 1              \\
                & \multicolumn{2}{l}{Is AGN}            & 2455     & 1238           \\ \hline
        \multicolumn{3}{l}{Is Weak Barred Disc and:}    &          &                \\
                & \multicolumn{2}{l}{Is Undetermined}   & 860      & 212            \\
                & \multicolumn{2}{l}{Is Uncertain}      & 4078     & 844            \\
                & \multicolumn{2}{l}{Is Star-forming}   & 19791    & 7903           \\
                & \multicolumn{2}{l}{Is Composite}      & 6545     & 2914           \\
                &           & LINER \oi                 & 595      & 391            \\
                &           & LINER \sii                & 1374     & 881            \\
                &           & LINER \nii                & 436      & 182            \\
                & \multicolumn{2}{l}{Is LINER}          & 2405     & 1454           \\
                &           & Optical AGN               & 2497     & 1048           \\
                &           & WISE AGN                  & 8        & 4              \\
                & \multicolumn{2}{l}{Is AGN}            & 2504     & 1051           \\ \hline
        \multicolumn{3}{l}{Is Strong Barred Disc and:}  &          &                \\
                & \multicolumn{2}{l}{Is Undetermined}   & 415      & 150            \\
                & \multicolumn{2}{l}{Is Uncertain}      & 1501     & 440            \\
                & \multicolumn{2}{l}{Is Star-forming}   & 4499     & 2075           \\
                & \multicolumn{2}{l}{Is Composite}      & 4158     & 2127           \\
                &           & LINER \oi                 & 579      & 437            \\
                &           & LINER \sii                & 1121     & 799            \\
                &           & LINER \nii                & 235      & 150            \\
                & \multicolumn{2}{l}{Is LINER}          & 1935     & 1386           \\
                &           & Optical AGN               & 1680     & 875            \\
                &           & WISE AGN                  & 1        & 0              \\
                & \multicolumn{2}{l}{Is AGN}            & 1680     & 875            \\ \hline \hline
        \end{tabular}
\end{table}

\section{Full stellar mass and colour distributions}\label{sec:full_dist}
\begin{figure}
    \centering
	\includegraphics[width=\columnwidth]{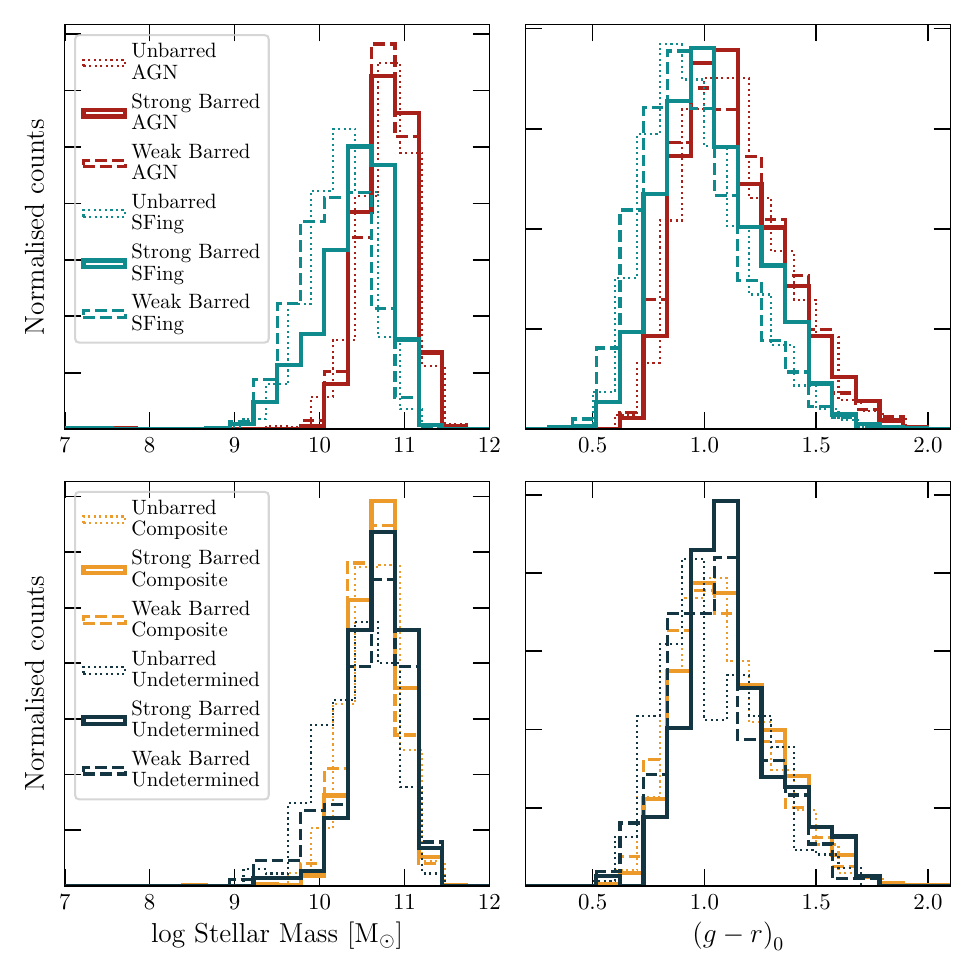}
    \caption{The distributions in stellar mass and $g-r$ colour for a variety of subsamples, with strongly barred galaxies in solid lines, unbarred in dotted lines and weakly barred in dashed lines. AGNs are in teal, star-forming in red, composite in orange and undetermined in navy blue.}
    \label{fig:agn_fix_mass_col_all}
\end{figure}

Fig. \ref{fig:agn_fix_mass_col_all} shows an identical plot to that in Fig. \ref{fig:agn_fix_mass_col}, with the addition of the distributions in mass and colour for weakly barred galaxies for completeness.

\bsp	
\label{lastpage}
\end{document}